\documentclass[12pt,a4paper]{article}
\usepackage{enumerate}
\usepackage{times}
\usepackage{a4wide}
\usepackage{amsfonts}
\usepackage{amssymb}
\usepackage{amsmath}
\usepackage{ifpdf}
\ifpdf
\usepackage[pdftex,unicode,implicit]{hyperref}
\hypersetup{%
  pdftitle    = {Lifshitz-like solutions with hyperscaling violation 
 in ungauged supergravity},
  pdfkeywords = {supergravity, duality, Lifshitz, String Theory},
  pdfauthor   = {Pablo Bueno, Wissam Chemissany, Patrick Meessen, Tom\'as
    Ort\'{\i}n and  C. S. Shahbazi},
  plainpages  = true,
  colorlinks  = true,
  citecolor   = blue,
  urlcolor    = red,
  linkcolor   = black
}
\newcommand{\hepth}[1]{{\tt
\href{http://www.arXiv.org/abs/hep-th/#1}{hep-th/#1}}}

\newcommand{\arxiv}[1]{{\tt
\href{http://www.arXiv.org/abs/#1}{#1}}}
\else
  \usepackage[dvips]{graphicx}
  \usepackage[unicode,implicit]{hyperref}
  \newcommand{\hepth}[1]{{\tt hep-th/#1}}

  \newcommand{\arxiv}[1]{{\tt #1}}
\fi
\usepackage{tikz}
\newcommand{\FPAUO}[2]{
\tikz[scale=.13,
         Uniovi/.style={color=green!51!blue, fill=green!51!blue}
 ] {
 \fill[Uniovi] (0,0) circle (10);
 \fill[white] (0,7) circle (1.5);
 \draw[Uniovi] (-2,7.5) rectangle (2,5.5);
 \fill[white] (-0.3,6.6) rectangle (0.3,0);   
 \fill[white] ( -0.9,6.2) rectangle (.9 ,5.6);
 \fill[white] (-1.4, 5.2) rectangle (1.4, 4.6);
 \fill[white] (0,0) ellipse (3.5 and 4);
 \fill[Uniovi] (-2.5,0.3) rectangle (2.5,-0.3);
 \fill[Uniovi] (-2,2.3) rectangle (2,1.7);
 \fill[Uniovi] (-2,-2.3) rectangle (2,-1.7);
 \fill[white] (-4.5,5.5) rectangle (-2.7,4.9);
 \fill[white] (-3.9,6.1) rectangle (-3.3,4.3);
 \fill[white] (4.5,5.5) rectangle (2.7,4.9);
 \fill[white] (3.9,6.1) rectangle (3.3,4.3);
 \foreach \x in { 0,..., 3 }
   \foreach \y in { 0,...,\x}
    {
     \fill[white] (-6-\x*0.7+\y*1.4,3.5-\x *1.97) -- (-5.6-\x*0.7+\y*1.4,2.4-\x *1.97) -- (-6.4-\x*0.7+\y*1.4,2.4-\x *1.97) -- cycle;
     \fill[white] (6-\x*0.7+\y*1.4,3.5-\x *1.97) -- (5.6-\x*0.7+\y*1.4,2.4-\x *1.97) -- (6.4-\x*0.7+\y*1.4,2.4-\x *1.97) -- cycle;
   };
 \draw (0,-6) node[
                               text centered, 
                               color=white, 
                               font={\fontsize{8}{4}\sffamily\selectfont}
                             ] {FPAUO-#1/#2};
}} 
\makeatletter
\@addtoreset{equation}{section}
\makeatother

\begin{document}

\begin{flushright}
\small
\FPAUO{12}{11}\\
IFT-UAM/CSIC-12-85\\
September 7\textsuperscript{th}, 2012\\
\normalsize
\end{flushright}

\begin{center}

\vspace{.5cm}

{\LARGE {\bf Lifshitz-like solutions with hyperscaling violation \\[5mm]
 in ungauged supergravity}}

\vspace{.7cm}

\begin{center}

\renewcommand{\thefootnote}{\alph{footnote}}
{\sl\large P.~Bueno$^{\heartsuit}$}
\footnote{E-mail: {\tt pab.bueno [at] estudiante.uam.es}},
{\sl\large W.~Chemissany$^{\clubsuit}$}
\footnote{E-mail: {\tt chemissany.wissam [at] gmail.com}},
{\sl\large P.~Meessen$^{\spadesuit}$}
\footnote{E-mail: {\tt meessenpatrick [at] uniovi.es}},
{\sl\large T.~Ort\'{\i}n$^{\heartsuit}$}
\footnote{E-mail: {\tt Tomas.Ortin [at] csic.es}}
{\sl\large and C.~S.~Shahbazi$^{\heartsuit, \diamondsuit}$}
\footnote{E-mail: {\tt Carlos.Shabazi [at] uam.es}}
\renewcommand{\thefootnote}{\arabic{footnote}}

\vspace{.4cm}

${}^{\heartsuit}${\it Instituto de F\'{\i}sica Te\'orica UAM/CSIC\\
C/ Nicol\'as Cabrera, 13--15,  C.U.~Cantoblanco, 28049 Madrid, Spain}\\

\vspace{.2cm}

${}^{\clubsuit}${\it Department of Physics and Astronomy, University of Waterloo, Waterloo,\\
Ontario, Canada, N2L 3G1}\\

\vspace{.2cm}

${}^{\spadesuit}${\it HEP Theory Group, Departamento de F\'{\i}sica, Universidad de Oviedo\\
  Avda.~Calvo Sotelo s/n, 33007 Oviedo, Spain}\\

\vspace{.2cm}

${}^{\diamondsuit}${\it Stanford Institute for Theoretical Physics 
and Department of Physics, Stanford University,\\
  Stanford, CA 94305-4060, USA}\\

\vspace{.5cm}

\end{center}

{\bf Abstract}

\begin{quotation}

  {\small 
    In this note we describe several procedures to construct, from known
    black-hole and black-brane solutions of any ungauged supergravity theory,
    non-trivial gravitational solutions whose ``near-horizon'' and
    ``near-singularity'' limits are Lifshitz-like spacetimes with dynamical
    critical exponent $z$,   ``hyperscaling violation'' exponent $\theta$
    and Lifshitz radius $\ell$ that depends on the physical parameters of the
    original black-hole solution.  Since the new Lifshitz-like solutions can
    be constructed from any black-hole solution of any ungauged supergravity,
    many of them can be easily embedded in String Theory. Some of the
    procedures produce supersymmetric Lifshitz-like solutions.
}

\end{quotation}

\end{center}

\setcounter{footnote}{0}

\newpage
\pagestyle{plain}

\tableofcontents

\section{Introduction}
\label{sec-Intro}

Gauge/gravity duality has found new and interesting applications in the
study of strongly coupled condensed matter systems \cite{Hartnoll:2009sz,
  Herzog:2009xv,McGreevy:2009xe}.  In this context one has to work with the
metrics that are dual to scale-covariant field theories which are not
conformally invariant. These theories are characterized by a dynamical
critical exponent $z\neq 1$ and a \textit{hyperscaling violation} exponent
$\theta \neq 0$ \cite{Charmousis:2010zz,Huijse:2011ef,Dong:2012se,
  Shaghoulian:2011aa}. The values $z=1$ and $\theta=0$ correspond to
conformally-invariant theories dual to the AdS metric. For other values of $z$
and $\theta$, in terms of dimensionless coordinates $t,r,x^{i}$, the
$(d+2)$-dimensional spacetime metric can be cast in the form
\begin{equation}
\label{eq:thetaz}
ds_{d+2}^{2} 
=
\ell^{2}  r^{-2(d-\theta)/d} 
\left[ 
r^{-2(z-1)} dt^{2} - dr^{2} - dx^{i}dx^{i}
\right]\, ,
\end{equation}
where $d$ is the number of spatial dimensions on which the dual theory lives
($i=1,\ldots ,d$) and the parameter $\ell$, with dimensions of length, is the
\textit{Lifshitz radius}.  We will refer, henceforth, to these metrics as
\textit{hyperscaling-violating Lifshitz (hvLf) metrics}.

hvLf geometries (\ref{eq:thetaz}) with the particular hyperscaling violation
exponent $\theta=d-1$ are intimately connected with compressible states with
hidden Fermi surfaces as well as with logarithmic violations of the area law
of entanglement entropy (see for instance
\cite{Huijse:2011ef,Dong:2012se,Shaghoulian:2011aa,Ogawa:2011bz}).  hvLf
solutions have also been of interest for their connection with string theory
and supergravity. We refer the reader to
\textit{e.g.}~\cite{Dong:2012se,Perlmutter:2012he,Narayan:2012hk,Ammon:2012je}.

It is known since the advent of the AdS/CFT duality that considering
temperature in the gauge theory corresponds to putting a black hole in the
bulk of the gravity side \cite{Witten:1998qj,Witten:1998zw}. When the gauge
theory is conformal, the corresponding black hole has AdS asymptotics in the
boundary. Similarly, since Lifshitz field theory with hyperscaling violation
has anisotropic scale covariance, the black holes describing the geometry dual
to its finite temperature generalization must have a metric of the form
(\ref{eq:thetaz}) as asymptotic geometry.

Asymptotically Lifshitz black holes with $\theta=0$ and $z\neq 1$ have been
extensively studied over the last few years. Analytic and numerical solutions
to gravitational theories with simple types of matter have been constructed.
String theoretic black hole solutions having Lifshitz asymptotics with
$\theta=0$ with a general dynamical exponent have been numerically computed
(see \cite{Amado:2011nd, Barclay:2012he} and references therein). Analytical
black hole solutions which could be related to string and supergravity
theories are still missing, though\footnote{The analytic black hole solutions
  in \cite{Chemissany:2011mb} suffer from naked singularities.}

So far, hvLf metrics (\ref{eq:thetaz}) with $\theta\neq 0$ have only been found in
solutions to Einstein-Maxwell-dilaton-type effective actions of the form
\begin{equation}
S
=
\frac{1}{16\pi G_{N}}
\int\sqrt{|g|}
\left\{
R
+\tfrac{1}{2}\partial_{\mu}\phi\partial^{\mu}\phi
-Z(\phi)F^{\mu\nu}F_{\mu\nu}
-2\Lambda
-V(\phi)
\right\}\, .
\end{equation}
Simple analytic black hole solutions have been constructed for this model for
specific choices of $V(\phi)$ and $Z(\phi)$ in
\cite{Huijse:2011ef,Dong:2012se,Shaghoulian:2011aa,Ogawa:2011bz,Iizuka:2011hg};
analytical hvLf solutions to a model with 2 gauge fields and an exponential scalar potential,
are presented and analysed in ref.~\cite{art:Eoin}. 
Finding embeddings of these models and solutions in gauged supergravities and,
eventually, in string theory would be most interesting, in particular for
asymptotically hvLf $\theta=d-1$ black holes.

In this work we report progress in this direction. In particular, we are going
to show how to construct systematically solutions of ungauged supergravity
whose metrics are, or approach in certain limits, hvLf metrics with certain
values of $z$ and $\theta$. The first of our constructions makes use of the
FGK formalism originally developed to study static, spherically symmetric,
asymptotically flat, black hole solutions of 4-dimensional ungauged
supergravity theories \cite{Ferrara:1997tw}, and we start by reviewing this
formalism in Section~\ref{sec-FGK}. We will then generalize the FGK formalism
to metrics which are not spherically symmetric. The main result is that there
are (at least) two cases in which the equations of motion of the metric
function and scalar fields are identical to those of the spherically symmetric
one. Thus, one can use the solutions of the standard black hole case and
construct solutions with entirely different spacetime metrics.

In section~\ref{sec-Lifs} we study the behaviour of the new solutions in the
neighborhood of the values of the radial coordinate corresponding, in the
original solution, to the inner and outer horizons, spatial infinity and the
curvature singularity. We will find hvLf metrics in some of these limits.  In
Section~\ref{sec-more} we investigate how hvLf metrics arise in other limits
of more standard metrics and propose other procedures to construct, in
particular, supersymmetric hvLf spacetimes by \textit{smearing} extremal
supersymmetric black hole solutions of $N=2,d=4$ supergravity. In
Section~\ref{sec-Lifs} we briefly discuss the generalization of these results
to higher dimensions. A brief discussion of our results can be found in
Section~\ref{sec-discussion} and the appendix contains a summary of properties
of hvLf metrics.


\section{The generalized FGK formalism }
\label{sec-FGK}


Following Ref.~\cite{Ferrara:1997tw} we consider the action
\begin{eqnarray}
\label{eq:generalaction}
I
\!=\!
 \int\!\! d^{4}x \sqrt{|g|}
\left\{
R +\mathcal{G}_{ij}(\phi)\partial_{\mu}\phi^{i}\partial^{\mu}\phi^{j} + 
+ 2 \Im{\rm m}\mathcal{N}_{\Lambda\Sigma}
F^{\Lambda}{}_{\mu\nu}F^{\Sigma\, \mu\nu}
-2 \Re{\rm e}\mathcal{N}_{\Lambda\Sigma}
F^{\Lambda}{}_{\mu\nu}\star F^{\Sigma\, \mu\nu}
\right\}\,  ,   
\end{eqnarray}
where $\mathcal{N}_{\Lambda\Sigma}$ is the complex scalar-dependent
(\textit{period}) matrix. The bosonic sector of any ungauged supergravity
theory in 4 dimensions can be put in this form. The number of scalars labeled
by $i,j,\cdots$ and of vector field labeled by $\Lambda, \Sigma,\cdots$, the
scalar metric $\mathcal{G}_{ij}$ and the period matrix
$\mathcal{N}_{\Lambda\Sigma}$ depend on the particular theory under
consideration.

Since we want to obtain static solutions, we consider the metric
\begin{equation}
\label{eq:generalbhmetric}
ds^{2} 
 = 
e^{2U} dt^{2} - e^{-2U} \gamma_{\underline{m}\underline{n}}
dx^{\underline{m}}dx^{\underline{n}}\, , 
\end{equation}
\noindent
where $\gamma_{\underline{m}\underline{n}}$ is a 3-dimensional
(\textit{transverse}) Riemannian metric to be specified later. Using
Eq.~(\ref{eq:generalbhmetric}) and the assumption of staticity of all the
fields, we perform a dimensional reduction over time in the equations of
motion that follow from the above general action. We obtain a set of reduced
equations of motion that we can write in the form\footnote{See
  Ref.~\cite{Ferrara:1997tw} for more details on this reduction.}
\begin{eqnarray}
\label{eq:Eq3dim1}
\nabla_{\underline{m}}
\left(\mathcal{G}_{AB} \partial^{\underline{m}}\tilde{\phi}^{B}\right)
-\tfrac{1}{2}\partial_{A} \mathcal{G}_{BC}
\partial_{\underline{m}}\tilde{\phi}^{B}\partial^{\underline{m}}\tilde{\phi}^{C} 
& = & 
0\, .
\\
& & \nonumber \\
\label{eq:Eq3dim2}
R_{\underline{m}\underline{n}}
+\mathcal{G}_{AB}\partial_{\underline{m}}\tilde{\phi}^{A}
\partial_{\underline{n}}\tilde{\phi}^{B} 
& = & 
0\, .
\\
& & \nonumber \\
\label{eq:Eq3dim3}
\partial_{[\underline{m}}\psi^{\Lambda}\partial_{\underline{n}]}\chi_{\Lambda} 
& = & 
0\, ,
\end{eqnarray}
where all the tensor quantities refer to the 3-dimensional metric
$\gamma_{\underline{m}\underline{n}}$ and we have defined the metric
$\mathcal{G}_{AB}$
\begin{equation}
\mathcal{G}_{AB}
\equiv
\left(
  \begin{array}{ccc}
   2 &  &  \\
  & \mathcal{G}_{ij} &  \\
  &   & 4 e^{-2U}\mathcal{M}_{MN} 
  \end{array}
\right)\, ,
\end{equation}
in the \emph{extended} manifold of coordinates
$\tilde{\phi}^{A}=\left(U,\phi^{i},\psi^{\Lambda},\chi_{\Lambda}\right)$,
where
\begin{equation}
(\mathcal{M}_{MN})
\equiv
\left( 
\begin{array}{lr}
    (\mathfrak{I}+\mathfrak{R}\mathfrak{I}^{-1}\mathfrak{R})_{\Lambda\Sigma} &
    -(\mathfrak{R}\mathfrak{I}^{-1})_{\Lambda}{}^{\Sigma} \\
    & \\
    -(\mathfrak{I}^{-1}\mathfrak{R})^{\Lambda}{}_{\Sigma} &
    (\mathfrak{I}^{-1})^{\Lambda\Sigma} \\   
  \end{array}
\right)
\, ,
\hspace{.3cm}
\mathfrak{R}_{\Lambda\Sigma} \equiv \Re{\rm e}\mathcal{N}_{\Lambda\Sigma}\, ,
\hspace{.3cm}
\mathfrak{I}_{\Lambda\Sigma} \equiv \Im{\rm m}\mathcal{N}_{\Lambda\Sigma}\, .
\end{equation}
Eqs.~(\ref{eq:Eq3dim1}) and (\ref{eq:Eq3dim2}) can be obtained from a
three-dimensional effective action
\begin{equation}
\label{eq:Eq3dim3action3dim}
I=\int\! d^{3}x \sqrt{|\gamma|}
\left\{ R 
+\mathcal{G}_{AB}\partial_{\underline{m}}\tilde{\phi}^{A}
\partial^{\underline{m}}\tilde{\phi}^{B}\right\}\, ,
\end{equation}
but we still need to add the constraint Eq.~(\ref{eq:Eq3dim3}).

If we now decide to consider spherically-symmetric transverse metrics only, as
it is appropriate to describe single, static black holes, we can choose, as
in Ref.~\cite{Ferrara:1997tw}
\begin{equation}
\label{eq:gammaBH}
\gamma_{\underline{m}\underline{n}}
dx^{\underline{m}}dx^{\underline{n}}
 = 
\frac{d\tau^{2}}{W_{-1}^{4}} 
+
\frac{d\Omega^{2}_{-1}}{W^{2}_{-1}}\, ,
\end{equation}
where $W_{-1}$ is a function of the (inverse) radial coordinate $\tau$ to be
determined and
\begin{equation}
d\Omega^{2}_{-1} \equiv d\theta^{2}+\sin^{2}{\! \theta}\, d\phi^{2}\, ,
\end{equation}
is the metric of the round 2-sphere of unit radius. With this choice,
Eq.~(\ref{eq:Eq3dim3}) is automatically solved, the equation of $W_{-1}(\tau)$
can be integrated completely, giving
\begin{equation}
\label{eq:Eq3tau}
W_{-1}(\tau)=\frac{\sinh\left(r_{0} \tau\right)}{r_{0}}\, ,
\end{equation}
and we are left with just
\begin{eqnarray}
\label{eq:Eq1tau}
\frac{d}{d\tau} \left(\mathcal{G}_{AB}
  \frac{d\tilde{\phi}^{B}}{d\tau}\right)
-\tfrac{1}{2}\partial_{A}\mathcal{G}_{BC}\frac{d\tilde{\phi}^{B}}{d\tau}
\frac{d\tilde{\phi}^{C}}{d\tau}
& = & 
0\, ,
\\
& & \nonumber \\
\label{eq:Eq2tau}
\mathcal{G}_{BC}\frac{d\tilde{\phi}^{B}}{d\tau}
\frac{d\tilde{\phi}^{C}}{d\tau}
& = & 
2 r^{2}_{0}\, .
\end{eqnarray}
The integration constant $r_{0}$ is the \textit{non-extremality parameter}:
when $r_{0}$ vanishes, the metric describes extremal black holes (if the
solution satisfies the necessary regularity conditions).

The electrostatic and magnetostatic potentials $\psi^{\Lambda},\chi_{\Lambda}$
only appear through their $\tau$-derivatives. The associated conserved
quantities are the magnetic and electric charges
$p^{\Lambda},q_{\Lambda}$ and can be used to eliminate completely the
potentials. The remaining equations of motion can be put in the convenient
form 
\begin{eqnarray}
\label{eq:e1}
U^{\prime\prime}
+e^{2U}V_{\rm bh}
& = & 0\, ,\\ 
& & \nonumber \\
\label{eq:Vbh-r0-real}
(U^{\prime})^{2} 
+\tfrac{1}{2}\mathcal{G}_{ij}\phi^{i\, \prime}  \phi^{j\, \prime}  
+e^{2U} V_{\rm bh}
& = & r_{0}^{2}\, ,\\
& & \nonumber \\
\label{eq:e3}
(\mathcal{G}_{ij}\phi^{j\, \prime})^{\prime}
-\tfrac{1}{2} \partial_{i}\mathcal{G}_{jk}\phi^{j\, \prime}\phi^{k\, \prime}
+e^{2U}\partial_{i}V_{\rm bh}
& = & 0\, ,
\end{eqnarray}
in which the primes indicate differentiation with respect to $\tau$ and the
so-called \textit{black-hole potential} $V_{\rm bh}$ is given by\footnote{As
  in Ref.~\cite{Galli:2011fq}, we adopt the sign of the black-hole potential
  opposite to most of the literature on black-hole attractors, conforming
  instead to the conventions of Lagrangian mechanics.}
\begin{equation}
-V_{\rm bh}(\phi,\mathcal{Q})
\equiv
-\tfrac{1}{2}\mathcal{Q}^{M}\mathcal{Q}^{N} \mathcal{M}_{MN}\, ,
\hspace{1cm}
(\mathcal{Q}^{M})
\equiv
\left(
  \begin{array}{c}
   p^{\Lambda} \\ q_{\Lambda} \\ 
  \end{array}
\right)\, .
\end{equation}
Eqs.~(\ref{eq:e1}) and (\ref{eq:e3}) can be derived from the effective action
\begin{equation}
\label{eq:effectiveaction}
I_{\rm eff}[U,\phi^{i}] = \int d\tau \left\{ 
(U^{\prime})^{2}  
+\tfrac{1}{2}\mathcal{G}_{ij}\phi^{i\, \prime}  \phi^{j\, \prime}  
-e^{2U} V_{\rm bh}
  \right\}\, .  
\end{equation}
Eq.~(\ref{eq:Vbh-r0-real}) is nothing but the conservation of the Hamiltonian
(due to absence of explicit $\tau$-dependence of the Lagrangian) but with a
particular value of the integration constant ($r_{0}^{2}$).

A fair number of solutions of this system for different theories of
$\mathcal{N}=2,d=4$ supergravity coupled to vector supermultiplets are known
(see \textit{e.g.}~Ref.~\cite{Galli:2011fq,Mohaupt:2011aa}). They describe
single, charged, static, spherically-symmetric, asymptotically-flat,
non-extremal black holes which generalize the Reissner-Nordstr\"om solution
and have two horizons that coincide when the non-extremality parameter $r_{0}$
vanishes. The metric covers the exterior of the outer (event) horizon when
the (\textit{inverse}) radial coordinate\footnote{Observe that $\tau$ has
  dimensions of inverse length, since $r_{0}$ has, conventionally, dimensions
  of length.} $\tau$ takes values in the interval $(-\infty, 0)$, whose limits
are, respectively, the event horizon and spatial infinity. The interior of the
inner (Cauchy) horizon corresponds to the interval $(\tau_{\rm s},+\infty)$,
whose limits are, respectively, the singularity and the inner horizon.

We may also be interested in spacetime metrics which are not spherically
symmetric, in which case we have to use a different transverse metric. In
principle, these metrics are not appropriate to describe isolated, static
black holes but here we are ultimately interested in Lifshitz metrics with a
transverse metric invariant under the 2-dimensional Euclidean group, Thus, we
can take, for instance, the following simple generalization of the
spherically-symmetric transverse metric Eq.~(\ref{eq:gammaBH}):
\begin{equation}
\label{eq:gammak}
\gamma_{\underline{m}\underline{n}}
dx^{\underline{m}}dx^{\underline{n}}
 = 
\frac{d\tau^{2}}{W_{\kappa}^{4}} 
+
\frac{d\Omega^{2}_{\kappa}}{W^{2}_{\kappa}}\, ,
\end{equation}
where $W_{\kappa}$ is a function of $\tau$ and $d\Omega^{2}_{\kappa}$ is the metric of
the 2-dimensional symmetric space of curvature $\kappa$ and unit radius:
\begin{eqnarray}
d\Omega^{2}_{-1} & \equiv & d\theta^{2}+\sin^{2}{\! \theta}\, d\phi^{2}\, ,
\\
& & \nonumber \\
d\Omega^{2}_{+1} & \equiv & d\theta^{2}+\sinh^{2}{\! \theta}\, d\phi^{2}\, ,
\\
& & \nonumber \\
\label{eq:domega0}
d\Omega^{2}_{0} & \equiv & d\theta^{2}+d\phi^{2}\, .
\end{eqnarray}
In these three cases the equation for $W_{\kappa}(\tau)$ can be integrated and the
results are 
\begin{eqnarray}
\label{eq:sinh}
W_{-1} 
& = & 
\frac{\sinh{r_{0} \tau}}{r_{0}}\, ,\\
& & \nonumber \\
\label{eq:cosh}
W_{1}
& = &
\frac{\cosh{r_{0} \tau}}{r_{0}}\, , \\
& & \nonumber \\
\label{eq:exp}
W^{\pm}_{0}
& = & 
a e^{\mp r_{0}\tau}\, ,
\end{eqnarray}
where $a$ is a real arbitrary constant with dimensions of inverse length. 

It turns out that if we follow now for the $\kappa=0,+1$ cases the procedure
described above for the $\kappa=-1$ case we arrive to exactly the same system
of equations (\ref{eq:e1})-(\ref{eq:e3}) and, therefore, to the same effective
action Eq.~(\ref{eq:effectiveaction}). It follows that all the solutions for
$\left(U,\phi^{i}\right)$ obtained in the spherically-symmetric case
$\kappa=-1$ are also solutions for the $\kappa=0,+1$ cases as well. In other
words: every solution of the system of equations~(\ref{eq:e1})-(\ref{eq:e3})
provides us with four different solutions of the original theory, by simply
using the four different transverse metrics.

Since, as mentioned above, there exists a number of solutions of those
equations that describe single, static, asymptotically-flat non-extremal black
holes when we take $\kappa=-1$
\cite{Galli:2011fq,Mohaupt:2011aa,Meessen:2011aa}, we can simply take those
solutions and study them setting $\kappa=0$ or $+1$ in the transverse
metric. Observe that one integration constant has been fixed to normalized the
metric at spatial infinity, something we may not need to do in the
$\kappa=0,+1$ cases, but the normalization could be changed at any moment, if
necessary.

In what follows we are going to study the asymptotic behaviour of generic
solutions $(U,\phi^{i})$, normalized to describe single, static,
asymptotically-flat non-extremal black holes for $\kappa=-1$ when we take the
transverse metric with $\kappa=0$\footnote{We leave the case $\kappa=+1$ for a
  future publication.}.


\section{Solutions with Lifshitz-like  asymptotics}
\label{sec-Lifs}


Since we are going to use the metric functions $e^{-2U}$ corresponding to
charged, spherically-symmetric, asymptotically-flat, non-extremal black-hole
solutions, we start by reviewing their asymptotic behaviors at the outer
($+$) and inner ($-$) horizons\footnote{Uncharged, static black holes only
  have outer horizon. The discussion of the behaviour of the metric function
  in the interior of the inner horizon does not apply to them.} (placed,
respectively, at $\tau=-\infty$ and $\tau=+\infty$) and at spatial infinity
$\tau=0$.
\begin{itemize}

\item The standard normalization of these asymptotically-flat black holes
  requires that 
\begin{equation}
\lim_{\tau\rightarrow 0^{-}}e^{-2 U}=1\, .
\end{equation}
\item When $\tau$ approaches the two horizons, $\tau\rightarrow \mp \infty$,
  the metric function behaves as
\begin{equation}
e^{-2U}\sim  \frac{S_{\pm}}{4\pi r_{0}^{2}}e^{\mp 2r_{0}\tau}\, ,   
\end{equation}
where $S_{+}$ (resp.~$S_{-}$) is the entropy of the outer (resp.~inner)
horizon, which is assumed to be non-vanishing (which is equivalent to require
regularity of the black-hole solution). If we use the spherically-symmetric
transverse metric the spacetime metric approaches in these limits a product of
a Rindler metric and a 2-sphere of area $4S_{\pm}$. Studying the Rindler
metric by conventional methods one finds that the temperatures of the horizons
$T_{\pm}$ obey the Smarr-like relation \cite{Gibbons:1996af}
\begin{equation}
\label{eq:smarr}
r_{0}=2S_{\pm}T_{\pm}\, .  
\end{equation}
\item $e^{-2U}$ vanishes for some value of $\tau_{\rm s}\in (0,+\infty)$ at
  which the physical singularity of the black-hole spacetime lies. We may also
  want to study the behaviour of $e^{-2U}$ near this value of $\tau$ but we do
  not know of any general result on this respect. We will have to study each
  particular case separately.
\end{itemize}

To find new solutions, we are going to plug black-hole metric functions in the
general static metric Eq.~(\ref{eq:generalbhmetric}) with the transverse
metric Eq.~(\ref{eq:gammak}) with $\kappa=0$, \textit{i.e.}~with
Eq.~(\ref{eq:domega0}) and Eq.~(\ref{eq:exp}). It is convenient to set
$a=1/r_{0}$ so no new length scale is introduced in the metric, which takes
two possible forms:
\begin{equation}
\label{eq:generalbhmetricexp}
ds^{2}_{(\pm)} 
= 
e^{2U} dt^{2} - e^{-2U}\left[  e^{\pm 4 r_{0}\tau} r^{4}_{0}d\tau^{2} 
+
e^{\pm 2 r_{0}\tau}r_{0}^{2}\left(d\theta^{2}+d\phi^{2} \right)  \right] \, .  
\end{equation}


\paragraph{Asymptotic behaviour of $ds^{2}_{(-)}$:}

 
Using the general properties of the metric function $e^{-2U}$ described above
it is easy to see that in the limit $\tau\rightarrow -\infty$ this metric
behaves as
\begin{equation}
\label{eq:Lifshitztau--}
ds^{2}_{(-)} 
\sim  
\frac{4\pi r_{0}^{2}}{S_{+}}e^{2r_{0}\tau} dt^{2} 
- \frac{S_{+}}{4\pi r_{0}^{2}}e^{-2r_{0}\tau} 
\left[  e^{-4 r_{0}\tau}r^{4}_{0}d\tau^{2} 
+
e^{-2 r_{0}\tau}r_{0}^{2} \left(d\theta^{2}+d\phi^{2} \right)\right] \, .  
\end{equation}
The change of coordinates
\begin{equation}
r \equiv e^{-r_{0}\tau}\, ,
\hspace{1cm}
\tilde{t} \equiv \frac{4\pi r_{0}^{2}}{S_{+}} t/r_{0}\, ,
\hspace{1cm}
x^{1} \equiv \theta\, ,
\hspace{1cm}
x^{2} \equiv \phi\, ,
\end{equation}
brings the metric to the form
\begin{equation}
\label{eq:Lifshitzr--}
ds^{2}_{(-)}
\sim
\frac{S_{+}}{4\pi} r^4
\left[
r^{-6} d\tilde{t}^{\, 2} -dr^{2} -dx^{i}dx^{i}
\right]\, ,
\end{equation}
which is a hvLf metric of the form Eq.~(\ref{eq:thetaz}) with $z=4,~\theta=6$ and
radius
\begin{equation}
\label{eq:ln}
\ell \sim r_{0}\, ,  
\end{equation}
up to dimensionless factors (functions of the quotient $S_{+}/r_{0}^{2}$);
observe that this asymptotic hvLf space lies in the class of Ricci flat
hvLf spaces in Eq.~(\ref{eq:RicciFlatHVLf}).

The metric $ds^{2}_{(-)}$ is regular at $\tau=0$. Spatial infinity is not
there because the radial distance between points with $\tau=0$ and points with
$\tau<0$ is finite and not infinite, as in the black-hole case. For $\tau$
equal to a certain $\tau_{\rm s}$, $e^{-2U}=0$ and the metric will be
singular, as in the black-hole case. Finally, in the $\tau\rightarrow +\infty$
limit the metric is the product of Rindler spacetime times $\mathbb{R}^{2}$,
which can be understood as a flat event horizon with the same temperature as
that of the inner horizon of the associated black-hole solution.


\paragraph{Asymptotic behaviour of $ds^{2}_{(+)}$:}


The analysis is completely analogous to the previous case: in the limit
$\tau\rightarrow -\infty$ we find a flat event horizon whose temperature is
that of the outer horizon of the associated black hole, there is a singularity
at $\tau=\tau_{\rm s}$ and a hyperscaling Lifshitz metric in the
$\tau\rightarrow +\infty$ limit. The Lifshitz radius is, once again,
$\ell=r_{0}$.


\subsection{Examples}
\label{sec-examples}

\paragraph{The Schwarzschild black hole:}


This is the only uncharged, static, spherically-symmetric, black-hole solution
of the class of theories we are considering and has only one horizon (the
event horizon) at (conventionally) $\tau\rightarrow -\infty$ in these
coordinates, which only cover the exterior. The metric function for the
Schwarzschild black hole in these coordinates is
\begin{equation}
\label{eq:metricfunctionSchwarzschild}
e^{-2U}= e^{2M\tau}\, ,
\end{equation}
The spacetime metric $ds^{2}_{(-)}$ constructed with the Schwarzschild metric
function takes the explicit form
\begin{equation}
ds^{2}_{(-)} 
= 
e^{-2M\tau} dt^{2} - e^{-2M\tau} M^{4}d\tau^{2} 
-
M^{2}\left(d\theta^{2}+d\phi^{2} \right)\, .    
\end{equation}
In the coordinates
\begin{equation}
e^{-M\tau}M \equiv r\, ,  
\end{equation}
it reads
\begin{equation}
ds^{2}_{(-)} 
= 
r^{2} d(t/M)^{2} - dr^{2} 
-
M^{2}\left(d\theta^{2}+d\phi^{2} \right)\, .    
\end{equation}
which is the product of a 2-dimensional Rindler spacetime ($\mathcal{R}i^{2}$)
with $\mathbb{R}^{2}$. The temperature of the flat horizon would be that of
the Schwarzschild black hole $T\sim M^{-1}$. Observe that this is not just the
asymptotic behaviour of the metric: the metric is everywhere identically
$\mathcal{R}i^{2}\times \mathbb{R}^{2}$. As is well-known, this metric is
just a wedge of the Minkowski spacetime which can be recovered by analytical
extension of this metric.

Observe that the above metric makes sense for $\tau\in (-\infty,+\infty)$
or $r\in (0,+\infty)$ since as discussed above, there is not spatial infinity
at $\tau=0$.

The metric $ds^{2}_{(+)}$ is in this case
\begin{equation}
ds^{2}_{(+)} 
= 
e^{-2M\tau} dt^{2} - e^{6M\tau} M^{4}d\tau^{2} 
-
e^{4M\tau} M^{2}\left(d\theta^{2}+d\phi^{2} \right)\, ,
\end{equation}
and, in the coordinates
\begin{equation}
e^{M\tau} \equiv r\, ,  
\end{equation}
it takes the form
\begin{equation}
ds^{2}_{(+)} 
= 
r^{-2} d(t/M)^{2} - r^{4} M^{2}dr^{2} 
-
r^{4} M^{2}\left(d\theta^{2}+d\phi^{2} \right)
=
M^{2}r^{4}
\left\{r^{-6} dt^{2} - dr^{2} 
-
d\theta^{2}-d\phi^{2}
\right\}
\, ,
\end{equation}
which is the $z=4,\theta=6, \ell\sim M$ hvLf metric everywhere in the
spacetime, and not just asymptotically. Yet again, this metric is defined for
all values of $\tau$ or for all $r\in (0,+\infty)$.


\paragraph{The Reissner-Nordstr\"om black hole:}


The embedding of the Reissner-Nordstr\"om black hole in pure
$\mathcal{N}=2,d=4$ supergravity (the supersymmetrization of the
Einstein-Maxwell theory). The metric function of this solution in the $\tau$
coordinates is \cite{Galli:2011fq}
\begin{equation}
e^{-2U} 
= 
\left[
\cosh{r_{0}\tau}
-\frac{M}{r_{0}}\sinh{r_{0}\tau}
\right]^{2}
\, ,
\hspace{1cm}
r^{2}_{0} \equiv M^{2} -|\mathcal{Z}|^{2}\, ,
\end{equation}
where 
\begin{equation}
\mathcal{Z} = \tfrac{1}{2}p -iq\, ,  
\end{equation}
is the central charge of pure $\mathcal{N}=2,d=4$ supergravity in the chosen
conventions.

It is evident that the asymptotic behaviour of the metrics $ds^{2}_{(\pm)}$
fits in the general case discussed above. 
Having the explicit form of the
metric, we can also study the behaviour of the spacetime metric near the
singularity at $\tau_{\rm s}$ at which $e^{-2U}(\tau_{\rm s})=0$. It is,
however, easier to do it in the coordinates in which the metric function has
the standard form
\begin{equation}
e^{-2U} = \frac{r^{2}}{(r-r_{+})(r-r_{-})}\, ,
\hspace{1cm}
r_{\pm} = M\pm r_{0}\, .  
\end{equation}
The coordinate transformation that relates these two forms of the metric
function is
\begin{equation}
r = -\left[\cosh{r_{0}\tau}
-\frac{M}{r_{0}}\sinh{r_{0}\tau} \right]
\left[ \frac{\sinh{r_{0}\tau}}{r_{0}} \right]^{-1}\, .
\end{equation}
If we make this coordinate transformation in the full $ds^{2}_{(\pm)}$
metrics, they take the form
\begin{equation}
\label{eq:deformedRN}
ds^{2}_{(\pm)}  
=
\frac{(r-r_{+})(r-r_{-})}{r^{2}} dt^{2} 
-\frac{r_{0}^{4}r^{2}}{(r-r_{\pm})(r-r_{\mp})^{5}}dr^{2}
-\frac{r_{0}^{2}r^{2}}{(r-r_{\mp})^{2}} (d\theta^{2} +d\phi^{2})\, .
\end{equation}

According to the general discussion, we should find the singularity in the
extension of the metric beyond $\tau=0$ to positive values of $\tau$. This
corresponds in these coordinates to values of $r$ ``beyond
$r=+\infty$''. Thus, we define the coordinate $u\equiv 1/r$ which overlaps
with $r$ for $u>0$ and extends the metric for $u\leq 0$. In these coordinates
the metric takes the form
\begin{equation}
ds^{2}_{(\pm)}  
=
(1-r_{+}u)(1-r_{-}u) dt^{2} 
-\frac{r_{0}^{4}}{(1-r_{\pm}u)(1-r_{\mp}u)^{5}} dr^{2}
-\frac{r_{0}^{2}}{(1-r_{\mp}u)^{2}} (d\theta^{2} +d\phi^{2})\, ,
\end{equation}
and, in the $u\rightarrow -\infty$ limit it approaches the metric
\begin{equation}
ds^{2}_{(\pm)}  
=
r_{+}r_{-}u^{2} dt^{2} 
-\frac{r_{0}^{4}}{r_{\pm}r_{\mp}^{5}u^{6}} du^{2}
-\frac{r_{0}^{2}}{r_{\mp}^{2}u^{2}} (d\theta^{2} +d\phi^{2})\, ,
\end{equation}
which can be put in the hvLf form with $z=3,\theta=4$ (which implies
$C(\theta,z)=0$) with the coordinate change $r^{\prime} \equiv 1/u$ using
rescaled the coordinates $\tilde{t}\equiv r_{\pm} t/r^{2}_{0}$, $\rho\equiv
r^{\prime}/r_{\mp}$, $x^{1}\equiv \sqrt{r_{+}r_{-}}/r_{0} \theta$,
$x^{2}\equiv \sqrt{r_{+}r_{-}}/r_{0} \phi$. 

Observe that the two consecutive coordinate changes $r=1/u$, $u=1/r^{\prime}$
mean that we can get the same result taking the limit of the metric when $r$
approaches $r=0$ (which corresponds to the value $\tau=\tau_{\rm s}$) ``from
the left''. In fact, the same result is obtained if we take the
near-singularity limit from the right.

Summarizing, the interior of the inner horizon region $r< r_{-}$ has,
therefore, two boundaries, at $r=r_{-}$ and at $r=0$. When the metric
approaches $r=r_{-}$ from the left, the metric $ds_{(+)}^{2}$ approaches a
hvLf metric with $z=4$ and $\theta=6$ and the metric $ds_{(-)}^{2}$ approaches
$\mathcal{R}i^{2}\times \mathbb{R}^{2}$, as we have seen before. When $r$
approaches $r=0$ the metric approaches a  hvLf metric with $z=3,\theta=4$.

The fact that a hvLf metric can describe the near-singularity limit of a
metric that has been obtained as a deformation of a regular black-hole metric
is very suggestive. Observe that the deformed metric Eq.~(\ref{eq:deformedRN})
differs from the standard Reissner-Nordstr\"om metric in factors of
$(r-r_{\pm})$, which are irrelevant in the $r\rightarrow 0$ limit, and in the
2-dimensional metric $d\Omega^{2}_{\kappa}$ which has $\kappa=-1$ for the
standard, spherically symmetric Reissner-Nordstr\"om black hole. In the next
section we are going to see that there is a limit of the Reissner-Nordstr\"om
black hole in which $d\Omega^{2}_{-1}$ approaches $d\Omega^{2}_{0}$. The
near-singularity limit of this Reissner-Nordstr\"om black hole will be
described by a hvLf metric with $z=3,\theta=4$.


\section{More hvLf metrics}
\label{sec-more}

In this subsection we want to discuss some other ways of obtaining hvLf spacetimes.


\subsection{hvLf spaces from limiting procedures}
\label{sec-limiting}

A 2-sphere looks locally (in small enough patches) like a plane. Thus, we can
flatten $d\Omega_{-1}^{2}$ by looking at a small neighborhood of
$\theta=\pi/2$ and we can study near-horizon and near-singularity limits of
standard, spherically-symmetric, black-hole solutions. The near-horizon limits
will give, obviously, $\mathcal{R}i^{2}\times \mathbb{R}^{2}$ metrics (or
$AdS_{2}\times\mathbb{R}^{2}$ metrics in the extremal cases). 

Let us consider the near-singularity limit of the Reissner-Nordstr\"om black
hole in a small patch around $\theta=\pi/2$:
\begin{eqnarray}
\label{eq:RN}
ds^{2} & =&
\frac{(r-r_{+})(r-r_{-})}{r^{2}} dt^{2} 
-\frac{r^{2}}{(r-r_{+})(r-r_{-})}dr^{2}
-r^{2} (d\theta^{2} + d\phi^{2}) \nonumber \\
 & & \nonumber \\
 & \sim &
\frac{r_{+}r_{-}}{r^{2}} dt^{2} 
-\frac{r^{2}}{r_{+}r_{-}}dr^{2}
-r^{2} (d\theta^{2} + d\phi^{2})\, ,
\end{eqnarray}
which can can be put in the hvLf form with $z=3$, $\theta=4$ and
$\ell=\sqrt{r_{+}r_{-}}$ with the coordinate change
$r/\sqrt{r_{+}r_{-}}\rightarrow r$, $t/\sqrt{r_{+}r_{-}} \rightarrow t$. 

We can also take the near-singularity limit of the Schwarzschild metric with
negative mass in a neighborhood  of $\theta=\pi/2$
\begin{eqnarray}
\label{eq:NegMassSchwarzschild}
ds^{2}
& =&
\left(1+\frac{2|M|}{r}\right)  dt^{2} 
-\left(1+\frac{2|M|}{r}\right)^{-1}dr^{2}
-r^{2} (d\theta^{2} + d\phi^{2}) \nonumber \\
 & & \nonumber \\
 & \sim &
\frac{2|M|}{r} dt^{2} 
-\frac{r}{2|M|}dr^{2}
-r^{2} (d\theta^{2} + d\phi^{2})\, ,
\end{eqnarray}
which can be put in the hvLf form with $z=4$, $\theta=6$ and $\ell= |M|/2$
with the coordinate change $2r/|M|\rightarrow \rho^{2}$, $4t/|M|\rightarrow
t$.


\subsection{Supersymmetric hvLf spaces from smearing}
\label{sec-smearing}


As was mentioned briefly in Section~\ref{sec-FGK}, the extremal limit
($r_{0}\rightarrow 0$) of the 4-dimensional metric describes a single static
black hole and the natural question, one we have been ignoring, is what
happens in the case $\kappa =0$.
\par
The first thing that changes is the asymptotic behaviour of $e^{-2U}$, which
for an extremal black hole reads
\begin{equation}
  \label{eq:10}
  \lim_{\tau\rightarrow -\infty}\ e^{-2U} \; =\; \frac{S}{\pi}\ \tau^{2} \; ,
\end{equation}
where $S$ is the entropy of the black hole. The second thing is that the
extremal limit of $W_{\kappa}^{\pm}$ is just the constant $a$ which has the
dimension of inverse length, whence the 4-dimensional metric becomes
\begin{equation}
\label{eq:3}
  ds^{2}_{0}\; =\; e^{2U}\ dt^{2}\ -\ 
e^{-2U}\ \left[ d(a^{-2}\tau )^{2} \ +\ d\vec{x}^{2}\ \right] \; ,
\end{equation}
where we have defined $x^{1}=\theta /a$ and $x^{2}=\phi /a$.  It is
straightforward to see that in the region $\tau\rightarrow -\infty$ this leads
to a hvLf space with $\theta =4$ and $z=3$. Similarly to what happens in the
Schwarzschild black hole case in Section~\ref{sec-examples}, one can see
that the extremal RN black hole of electrical charge $q$, which has $e^{-U}=
1-\frac{|q|}{\sqrt{2}}\tau$, is this asymptotic hvLf.

Now we are going to see that this solution is just a particular case of a very
wide class of solutions with hvLf asymptotics.

One of the most interesting features of the extremal RN black hole is that it
is supersymmetric in pure $N=2$, $d=4$ supergravity. As is well known, the
most general supersymmetric static solution of this theory can be written,
using Cartesian coordinates in the transverse space $\vec{y}_{3} \equiv 
(y^{1},y^{2},y^{3})$ as
\begin{equation}
  \label{eq:11}
  ds^{2}_{susy}\; =\; e^{2U}dt^{2} \ -\ e^{-2U}\ d\vec{y}_{3}^{\, 2} \; ,
\end{equation}
where the metric function has the form\footnote{We use the conventions of
  Ref.~\cite{Galli:2011fq}.}
\begin{equation}
e^{-2U}=\tfrac{1}{2}(H^{0})^{2}+2(H_{0})^{2}\, ,
\end{equation}
where $H_{0}$ and $H^{0}$ are two real harmonic functions in the flat
transverse space which satisfy the staticity constraint
\begin{equation}
\label{eq:staticity}
H^{0}\partial_{m} H_{0}-H_{0}\partial_{m} H^{0}=0\, ,
m=1,2,3\, . 
\end{equation}
In these coordinates, the standard, spherically symmetric ($\kappa=-1$),
purely electric extremal RN black hole corresponds to the choice of harmonic
functions
\begin{equation}
H^{0}=0\, ,
\hspace{1cm}
H_{0}= 1+\tfrac{1}{\sqrt{2}}\frac{|q|}{|\vec{y}_{3}|}\, .
\end{equation}

However, other choices (usually discarded when one is only interested in black
holes) are possible and are also supersymmetric. For instance, one can consider
harmonic functions that depend on only one of the transverse coordinates, say
$y^{3}\equiv \rho$. This corresponds, physically, to the smearing of the
spherically-symmetric solution in the $(y^{1},y^{2})$ plane and,
mathematically, to the substitution of the factor $1/r$ by $\rho$ in all the
harmonic functions of the spherically-symmetric solution. The staticity
constraint Eq.~(\ref{eq:staticity}) is automatically satisfied is it was in
the spherically-symmetric solution.

From the the extremal RN solution, this choice gives the new \textit{smeared}
solution
\begin{equation}
 ds^{2} = \tfrac{1}{2}(H_{0})^{-2} dt^{2} 
- 2(H_{0})^{2}\ \left[d\rho^{2} +dy^{i}dy^{i} \right] \; ,
\hspace{1cm}
H_{0}= 1+\tfrac{1}{\sqrt{2}}|q| \rho\, ,
\end{equation}
and this solution is identical to the $\kappa=0$ solution in
Eq.~(\ref{eq:3})\footnote{Observe that, in the non-extremal case, we cannot
  view the $\kappa=0$ solutions as the smearing of $\kappa=-1$ solutions.}
with $\tau=-\rho$. Furthermore, the $z\rightarrow \infty$ limit, which gives
the $\theta =3$, $z=4$ hvLf space corresponds to the choice
\begin{equation}
H_{0}= \tfrac{1}{\sqrt{2}}|q|\rho\, ,
\end{equation}
and, therefore, it is an exact, supersymmetric solution.

Once this connection between hvLf metrics and smeared supersymmetric black
holes of $N=2,d=4$ supergravity has been established, we can systematically
construct supersymmetric hvLf metrics using the well-known systematic
procedure to construct all the supersymmetric black hole solutions of any
$N=2,d=4$ supergravity coupled to vector supermultiplets
\cite{Behrndt:1997ny,LopesCardoso:2000qm,Denef:2000nb,Meessen:2006tu} and
choosing harmonic functions that depend on only one coordinate in transverse
space. The $\rho\rightarrow \infty$ limit is the same in all the cases (namely
a $\theta =3$, $z=4$ hvLf spacetime), provided that the original,
spherically-symmetric solution has a regular near-horizon limit. The scalar
fields, which have non-trivial profiles in the smeared solutions, become
constant in the $\rho \rightarrow \infty$ limits, just as they do in the
black-hole near-horizon limits.

There are, however, more possibilities, if we start from supersymmetric black
holes with singular horizon. A good example is provided by the supersymmetric
D0-D4 black holes embedded in the $STU$ model
\cite{Duff:1995sm,Behrndt:1996hu,Bellucci:2008sv}\footnote{Again, we use the
  notation and conventions of Ref.~\cite{Galli:2011fq} where the details can
  be found.}. After the smearing, the three complex scalars $Z^{i}$, $i=1,2,3$
and metric function of these solutions are given by
\begin{align}
\label{eq:STU1}
Z^{i} & = 
-4ie^{2U}H_{0}H^{i}\, ,
\\
&  \nonumber \\
\label{eq:STU2}
e^{-2U}
& =  
4\, \sqrt{
H_{0}H^{1}H^{2}H^{3}
}\, ,   
\end{align}
where the four harmonic functions $H_{0},H^{1},H^{2},H^{3}$ 
are
\begin{equation}
\begin{array}{rcl}
H_{0}
& = & 
s_{0}\, 
\left\{a_{0}
+\frac{1}{\sqrt{2}}{\displaystyle\frac{|q_{0}|}{|\vec{y}_{3}|}}\right\}\, ,
\\
 & & \\    
H^{i}
& = & 
s^{(i)}\, 
\left\{a^{(i)}
+\frac{1}{\sqrt{2}}{\displaystyle\frac{|p^{(i)}|}{|\vec{y}_{3}|}}\right\}\, ,
\\
\end{array}
\end{equation}
where $a_{0},a^{i}$ are constants related to the asymptotic ($r\rightarrow
\infty$) values of the scalars, $q_{0},p^{i}$ are electric and magnetic
charges and $s_{0},s^{i}$ are the signs of those charges. Only two sets of
signs of charges lead to supersymmetric and regular black holes: all charges
positive or negative.  In particular, none of these charges can vanish.

The associated smeared solutions are given by Eqs.~(\ref{eq:STU1}) and
(\ref{eq:STU2}) with the harmonic functions given by 
\begin{equation}
\begin{array}{rcl}
H_{0}
& = & 
s_{0}\, 
\left\{a_{0}
+b_{0}\rho\right\}\, ,
\\
 & & \\    
H^{i}
& = & 
s^{(i)}\, 
\left\{a^{(i)}
+b^{(i)}\rho\right\}\, .
\\
\end{array}
\end{equation}
The constants $b_{0},b^{i}$, which we can take to be positive, are related to
electric and magnetic fluxes. The staticity condition Eq.~(\ref{eq:staticity})
is satisfied for any values of the constants and, in particular, we can take
any number of them to vanish.

When all the $b_{0},b^{i}$ constants are different from zero, we can take all
the $a_{0},a^{i}$ to vanish or take the $\rho \rightarrow \infty$ limit. In
both cases $e^{-2U}= 4 \sqrt{ b_{0}b^{1}b^{2}b^{3} } \rho^{2}$ and we get a
$\theta =3$, $z=4$ hvLf spacetime with constant scalars.

When one of them ($b_{0}$, for instance) vanishes we must keep $a_{0}\neq 0$,
and we get  
\begin{equation}
Z^{i} = -i \frac{a_{0}b^{i}}{\sqrt{ a_{0}b^{1}b^{2}b^{3} }} \rho^{-1/2}\, ,
\hspace{1cm}
e^{-2U}= 4 \sqrt{ a_{0}b^{1}b^{2}b^{3} } \rho^{7/2}
\end{equation}
which is a $\theta =7/2$, $z=5/2$ hvLf spacetime, now with non-trivial
scalars. Other choices of vanishing constant $b$ lead to different scalar
profiles by the same $\theta$ and $z$.

It is easy to see that, for $n=0,\cdots, 4$ non-vanishing constants $b$, one
gets a hvLf spacetime with $\theta= 2+n/2$ and $z= 1+n/2$ and various scalar
profiles. Perhaps not surprisingly $C_{(\theta,z)}= (4-n)/n$ and only vanishes
for $n=4$.


\subsection{Higher dimensional generalization}
\label{sec-higher}

In ref.~\cite{Meessen:2011bd} the FGK formalism was Generalised to higher
dimensional cases, and it is only natural to consider the higher dimensional
generalizations of the results presented in the foregoing sections, starting
off by the ones in Section~\ref{sec-FGK}: the $D$-dimensional generalization
of the FGK metric reads
\begin{equation}
  \label{eq:1}
  ds^{2} = e^{2U}dt^{2} - e^{-\frac{2}{d-1}U}\ \left[
              \frac{d\rho^{2}}{(d-1)^{2}\ W_{\kappa}^{2d/(d-1)}} 
              \ +\
              \frac{\mathsf{h}_{ij}\ dx^{i}dx^{j}}{W_{\kappa}^{2/(d-1)}}
      \right] \; ,
\end{equation}
where $\mathsf{h}$ is the metric of a $d$-dimensional Riemannian Einstein
space; this metric is normalized such that
\begin{equation}
  \label{eq:9}
  R\left( \mathsf{h}\right)_{ij}\; =\; (d-1)\ \kappa\ \mathsf{h}_{ij} \; .
\end{equation}
The normalization is such that a $d$-sphere with the round metric has $\kappa =-1$.
\par
A so-so calculation then shows that the conditions for the resulting FGK
equations of motion to be $\kappa$ as well as $W_{\kappa}$ independent, are
\begin{equation}
  \label{eq:4}
  W_{\kappa}\ \ddot{W}_{\kappa}\ -\ \dot{W}_{\kappa}^{2}\; =\; \kappa 
  \hspace{.6cm}\mbox{and}\hspace{.6cm}
  \ddot{W}_{\kappa} \; =\; \mathcal{B}^{2}\ W_{\kappa} \; ;
\end{equation}
$\mathcal{B}$ plays the r\^{o}le of the $D$-dimensional non-extremality
constant which on dimensional grounds can be written as $r_{0}^{d-1}$.  The
solutions to the conditions (\ref{eq:4}) are
\begin{equation}
  \label{eq:5}
  W_{-1}\ =\ \frac{\sinh\left( \mathcal{B}\rho\right)}{\mathcal{B}} \;\;\; ,\;\;\;
  W_{0}^{\pm}\ =\ a\ e^{\mp \mathcal{B}\rho} 
  \hspace{.5cm}\mbox{and}\hspace{.5cm}
  W_{1}\ =\ \frac{\cosh\left(\mathcal{B}\rho\right)}{\mathcal{B}} \; .
\end{equation}
By looking at the, in general, fractional powers of $W$ that appear in the
metric (\ref{eq:1}), we see that in contradistinction to the 4-dimensional
case, the putative horizon lies at $\rho\rightarrow\infty$ which means that
the near-horizon behaviour for a non-extremal black hole implies
\begin{equation}
  \label{eq:2}
  \lim_{\rho\rightarrow\infty} e^{U} \; \sim\; e^{-\mathcal{B}\rho} \; .  
\end{equation}
The above means that given a solution to a $D$-dimensional FGK system, we can
as before deform the $\kappa =-1$ solution as in Section~\ref{sec-FGK}, and
obtain new solutions with properties similar to the ones encountered in the
foregoing sections. For example, concerning the $\rho\rightarrow\infty$
behaviour of the metric we see that


\paragraph{The $W^{+}_{0}$ case:} In this case the $\rho\rightarrow\infty$ spacetime
is hvLf with
\begin{equation}
  \label{eq:6}
  \theta\; =\;  \frac{d(d+1)}{d-1}  
  \hspace{.5cm}\mbox{and}\hspace{.5cm}
  z\; =\;   \frac{2d}{d-1}  \; ,
\end{equation}
which, as one can see from Eq.~(\ref{eq:RicciFlatHVLf}), corresponds to the
Ricci flat hvLf spaces.


\paragraph{The $W^{-}_{0}$ case:} Together with the condition (\ref{eq:2}), we
see that the resulting $\rho\rightarrow\infty$ spacetime is a Rindler wedge
times $\mathbb{R}^{d}$.

In the $\kappa =-1$ case, the validity of the $\rho$-coordinate, {\em i.e.\/}
$\rho\in [0,\infty )$ is principally determined by $W_{-1}$ and one imposes
conditions on $e^{U}$ in order to obtain metrics describing the spacetime
outside the outer horizon. In particular, the zero of $W_{-1}$ at $\rho =0$,
together with the regularity of $e^{U}$ there, allows for the identification
of $\rho =0$ with asymptotic spacetime.  $W_{0}$ is, however, an all-together
different beast and the naive validity of the coordinate, {\em i.e.\/}
$\rho\in [0,\infty )$, can be extended till one encounters a zero or a pole in
$e^{U}$; the former signaling a horizon, the latter a singularity.  Let us
illustrate this point with


\paragraph{The $5$-dimensional STU model:}

The FGK equations for the 5-dimensional STU model are completely separable,
whence the full analytical solution is known. The general solution satisfying
Eq.~(\ref{eq:2}) and having constant scalars in the limit
$\rho\rightarrow\infty$ is given by (see {\em e.g.\/}
\cite{Mohaupt:2011aa,Meessen:2012su})
\begin{equation}
  \label{eq:7}
  e^{-3U} \; =\; \frac{|\mathrm{q}_{1}\mathrm{q}_{2}\mathrm{q}_{3}|}{\mathcal{B}^{3}}\ 
                          \sinh\left( \alpha_{1}+\mathcal{B}\rho\right)\
                          \sinh\left( \alpha_{2}+\mathcal{B}\rho\right)\
                          \sinh\left( \alpha_{3}+\mathcal{B}\rho\right)\; ,
\end{equation}
where the $\mathrm{q}$ are the electrical charges and the $\alpha$'s are some
real constants; in the $\kappa =-1$ case they are chosen such that $U(\rho
=0)=1$ and one obtains a Minkowski space with the regular normalization. In
the $\kappa =0$ case, however, the point $\rho =0$ is not asymptotic and there
is therefore no need to impose said condition on the $\alpha$'s.  In fact, let
$0<\alpha_{1}\leq \alpha_{2}\leq \alpha_{3}$, then we can extend the
definition of $\rho$ to the point $\rho_{s} = -\alpha_{1}/\mathcal{B}$, where
we have added a subscript to highlight the fact that at that point we're
facing a curvature singularity.
\par
As in Section~\ref{sec-examples} we can consider the near-singularity
metric: in general we will find a hvLf space and the characteristic parameters
$(\theta ,z)$ will depend on the order of the zero of $e^{-3U}$ in
Eq.~(\ref{eq:7}). Denoting this number by $\mathsf{p}$, whence
$\mathsf{p}=1,2$ or $3$,\footnote{ To wit: $\mathsf{p}=1$ implies
  $\alpha_{2}>\alpha_{1}$, $\mathsf{p}=2$ means
  $\alpha_{3}>\alpha_{2}=\alpha_{1}$ and $\mathsf{p}=3$ means
  $\alpha_{3}=\alpha_{2}=\alpha_{1}$.  Let us in passing observe that the case
  $\mathsf{p}=3$ corresponds to the deformation of the 5-dimensional
  Reissner-Nordstr\"om black hole.  } we see that the near-singularity hvLf is
given by
\begin{equation}
  \label{eq:8}
  \theta\ =\ 3 + \frac{\mathsf{p}}{2} \;\; ,\;\;
  z\; =\; 1+ \frac{\mathsf{\mathsf{p}}}{2} \hspace{.5cm}\mbox{whence}\;
  C_{(\theta ,z)} \; =\; \frac{2(3-\mathsf{p})}{\mathsf{p}} \; \geq\; 0 \; ,
\end{equation}
and, furthermore, the null energy condition (\ref{eq:NullEnergy}) is always
satisfied.

In higher dimensions we can also construct hvLf solutions by smearing
extremal, supersymmetric black-hole solutions. The procedure is entirely
similar to the one followed in four dimensions. 

In a higher-dimensional context, it is natural to consider the following
brane-like generalization of the hvLf metric (\ref{eq:thetaz})
\begin{equation}
\label{eq:thetazbrane}
ds_{d+2}^{2} 
=
\ell^{2}  r^{-2(d-\theta)/d} 
\left[ 
r^{-2(z-1)} \left(dt^{2}- dy^{a}dy^{a}\right) - dr^{2} - dx^{i}dx^{i}
\right]\, ,
\hspace{.5cm}
a=1,\cdots,p\, ,
\,\,\,
i=1,\cdots d\, .
\end{equation}
The $p=0$ case is the original hvLf metric and a metric with $d=0$ and $p=\neq
0$ can be rewritten as a $p=0,d\neq 0$ by a coordinate change.

It should come as no surprise that we can obtain metrics of this kind by
smearing extremal supersymmetric $p$-brane metrics. As an example, consider
the 10-dimensional D$p$-brane solutions in the Einstein frame
\begin{equation}
  \begin{array}{rcl}
ds^{2} 
& =  &
H^{\frac{p-7}{8}}[dt^{2} -d\vec{y}^{\, 2}_{p}]
-H^{\frac{p+1}{8}} d\vec{x}^{\, 2}_{8-p}\, ,
\\
& & \\
C_{(p+1)\, t y^{1}\cdots y^{p}}
& = & 
\pm e^{-\phi_{0}}
(H^{-1} -1)\, ,
\\
& & \\
e^{-2\phi}
& = & 
e^{-2\phi_{0}} H^{\frac{p-3}{2}}\, .
\\
\end{array}
\end{equation}
In all cases, we can take\footnote{In the $p=8$ case there is no smearing
  involved, since there is only one transverse dimension.}
\begin{equation}
H \sim \rho\, ,  
\end{equation}
and put the metric in the form 
\begin{equation}
  \begin{array}{rcl}
ds^{2} 
& \sim  &
\rho^{\frac{p+1}{8}}\left\{ \rho^{-1}
[dt^{2} -d\vec{y}^{\, 2}_{p}]
-d\rho^{2}- d\vec{x}^{\, 2}_{8-p}\right\}\, ,
\\
& & \\
C_{(p+1)\, t y^{1}\cdots y^{p}}
& = & 
\sim\rho^{-1}\, ,
\\
& & \\
e^{-2\phi}
& \sim & 
 \rho^{\frac{p-3}{2}}\, ,
\\
\end{array}
\end{equation}
which is of the above form with $p=p$, $z=3/2$ and $\theta= (8-p)(p+17)/16$
for $p<8$. The case $p=0$ (the D0-brane) is a standard hvLf metric with $d=8$,
$\theta =8.5$ and $z=3/2$, which satisfies the null energy condition
(\ref{eq:NullEnergy}) but does not avoid the null curvature singularity in the
IR region ($\rho\rightarrow\infty$).  The string coupling constant reads
$e^{\phi}=r^{3/4}$, which goes to zero in the UV.  The case $p=8$, after a
change of coordinates $\varrho\equiv \rho^{3/2}$ is also a standard hvLf
metric ($p=0$) with $d=8$, $\theta=25/3$ and $z=1$ which also satisfies the
null energy condition (\ref{eq:NullEnergy}) but is singular in the IR region
($r\rightarrow\infty$). 

\section{Discussion}
\label{sec-discussion}


In this paper we have shown that hvLf metrics appear in many near-horizon and
near-singularity limits of well-known solutions or solutions that one can
easily construct by deforming them.  The abundance of examples seems to
suggest that hvLf metrics capture the behavior of many spacetimes near certain
curvature singularities; in particular near timelike singularities such as
those of the extremal RN black hole or the Schwarzschild solution with
negative mass.

Since some hvLf are holographically related to some well-known QFTs, this
apparently general property suggests the very attractive possibility of finding
holographically related QFTs that can describe those classical curvature
singularities, at least in some regime. Finding quantum systems with the right
values of $z$ and $\theta$ may be difficult, or impossible, though.  More
work is needed to see if this possibility can be realized.

\section*{Acknowledgments}
The authors would like to thank E. \'O Colg\'ain and S. Kachru for stimulating discussions.  WC and CSS
would like to thank the Stanford Institute for Theoretical Physics for
its hospitality; likewise, PM would like to thank the Instituto de F\'{\i}sica Te\'orica
its hospitality.  This work has been supported in part by the Spanish Ministry
of Science and Education grant FPA2009-07692, the Princip\'au d'Asturies grant
IB09-069, the Comunidad de Madrid grant HEPHACOS S2009ESP-1473, and the
Spanish Consolider-Ingenio 2010 program CPAN CSD2007-00042. The work of PM has
been supported by the Ram\'on y Cajal fellowship RYC-2009-05014.  The work of
PB and CSS has been supported by the JAE-predoc grants JAEPre 2011 00452 and
JAEPre 2010 00613. TO wishes to thank M.M.~Fern\'andez for her constant
support.


\appendix

\section{Some properties of the hvLf metrics}
\label{app:properties}
%
The hvLf metric (\ref{eq:thetaz})
is spatially homogeneous and covariant under the scale transformations
\begin{equation}
\label{scalecovariance}
x_{i} \to \lambda x_{i}
\;,\;
t \to \lambda^{z} t
\;,\;
r \to \lambda r
\;,\;
ds_{d+2}^{2} \to \lambda^{2\theta/d} ds_{d+2}^{2}\, ,
\end{equation}
where $\lambda$ is a dimensionless parameter. Observe that this means that the
Lifshitz radius $\ell$ is only defined up to dimensionless factors. The Ricci tensors of metrics (\ref{eq:thetaz}) are given by
\begin{eqnarray}
R_{tt}&=&\frac{(d z-\theta)(d+z-\theta)}{d\, r^{2z}}\\
R_{rr}&=&\frac{(d+z)\theta-d(z^2+d)}{d\, r^{2}}\\
R_{ij}&=& \frac{(\theta-d)(d+z-\theta)}{d\, r^{2}}\delta_{ij}.
\end{eqnarray}

This geometry generically suffers from a null curvature singularity at $r=\infty$
except for a specific set of parameter values. The singularity exists even
though all curvature invariants remain finite. The tidal forces diverge as 
\cite{Shaghoulian:2011aa}
\begin{equation} 
C_{(\theta,z)}r^{2 C_{(\theta,z)}+d}\, ,
\hspace{1cm}
C_{(\theta,z)}=\frac{d(z-1)-\theta}{d-\theta}
\end{equation}
where we have restricted to $C_{(\theta,z)}>0$ for which the singularity is a null curvature singularity as surfaces of constant $r$ become null as $r$ goes to infinity.
We distinguish several cases:
\begin{itemize}
\item For $\theta=0$ we simply get the result in \cite{Horowitz:2011gh} which
  is appropriate for Lifshitz scaling.  Ways for resolving the null curvature singularities
  have been presented in \cite{Harrison:2012vy, Bao:2012yt}.
\item The case of $\theta=0$ and $z=1$ is the non-singular result of pure AdS.
\item  There are non-singular results for  
\begin{equation} 
C_{(\theta,z)}=0\, ,
\,\,\,\,\textrm{or}\,\,\,\,\, 
C_{(\theta,z)}+1\leq 0\, .
\end{equation}
\end{itemize}
The null energy condition in the bulk gives the conditions
\begin{equation} 
\label{eq:NullEnergy} 
C_{(\theta,z)}\geq 0,\qquad (z-1) (d+z-\theta)\geq 0,
\end{equation}
which rules out the non-singular condition $C_{(\theta,z)}+1\leq 0$ and leaves
the condition $C_{(\theta,z)}=0$.
\par
There is a class of Ricci-flat hvLf spaces: they are characterized by 
\begin{equation}
\label{eq:RicciFlatHVLf}
 \theta\; =\;  \frac{d(d+1)}{d-1}  
  \hspace{.5cm}\mbox{and}\hspace{.5cm}
  z\; =\;   \frac{2d}{d-1}   \;\;\;\longrightarrow\;\; C_{(\theta ,z)}\; =\; 0 \; .
\end{equation}
These spaces always solve the null energy condition and are regular in the IR interior ($r\rightarrow \infty).$

\end{document}